# Beyond the Metric: Geometrical Measurability as a Constraint on Quantum Gravity

Matteo Tuveri[1,2,*]

[1] Physics Department, University of Cagliari, Cittadella Universitaria di Monserrato, 09042, Monserrato (CA), Italy
[2] INFN Sezione di Cagliari, University of Cagliari, Cittadella Universitaria, 09042, Monserrato (CA), Italy

*Corresponding author: matteo.tuveri@unica.it

**Abstract**

This paper develops an epistemological constraint on quantum gravity grounded in the empirical meaning of general relativity. The central claim is that a complete recovery of general relativity requires an effective metric, a continuum limit, or Einstein-like dynamics together with the physical conditions under which relational geometrical quantities can be objectively determined. These conditions concern the dynamical stability of measuring devices and reference systems, causal accessibility among physical systems, record formation, and invariance under admissible descriptions. In classical general relativity, they are usually implicit in the use of clocks, rods, light signals, freely falling bodies, detectors, and gauge-invariant observables. In quantum gravity, however, they become non-trivial because spacetime geometry may be emergent, effective, thermodynamic, relational, or frame-dependent. This claim is developed through four cases: Rindler horizons and the Unruh effect, black-hole thermodynamics and Jacobson's equation-of-state derivation, gravitational-wave detection, and Weyl and conformal gravity. The latter is discussed as a critical limiting case in which conformal invariance raises a sharp question about whether scale-dependent measurements of space and time can be physically fixed. Implications for quantum gravity are also discussed using emergent gravity and quantum reference frames as examples. The perspective developed in the study suggests a general epistemological constraint on quantum gravity: any viable approach must recover the physical possibility of objective geometrical measurement together with geometry itself.

## 1. Introduction

General relativity is commonly described as the theory in which gravitation is geometry. More precisely, gravitation is represented by a dynamical pseudo-Riemannian geometry with Lorentzian signature, whose metric determines proper time, causal structure, geodesic motion, curvature, gravitational redshift, light propagation, and the coupling between matter and gravitational dynamics [1,2]. This formulation has extraordinary empirical reach: it accounts for light bending, perihelion precession, gravitational time dilation, gravitational lensing, binary-pulsar energy loss, black-hole phenomenology, and gravitational-wave detection [1,3,4].

The slogan that gravitation is geometry can be interpreted in different ways, especially from the perspective of quantum gravity. On a strong geometrical reading, spacetime geometry is a fundamental physical structure whose properties are revealed by material systems and detector records. This view is naturally connected with approaches that seek to formulate a quantum theory of geometry or of quantum gravitational degrees of freedom [5-7]. On a more cautious reading, the geometrical structure of general relativity is the most compact and invariant representation of relations among physical systems: clocks, rods, light rays, freely falling bodies, detectors, and reference frames [8-12]. The distinction matters because emergent, effective, relational, thermodynamic, or postquantum approaches to gravity must explain how spacetime geometry arises and how such geometry becomes empirically coherent and functionally adequate to the production of evidence [13-16].

The standard empirical use of general relativity presupposes clocks, rods, light signals, freely falling bodies, detectors, records, and gauge-invariant or relational quantities. When these elements are properly defined and available, they may appear implicit and almost unproblematic within the classical theory. In quantum gravity, however, this is no longer automatic. The classical infrastructure of geometrical measurement cannot simply be presupposed; it must be recovered, derived, or otherwise justified in the appropriate limit.

This paper formulates this requirement as a general epistemological constraint on quantum gravity. Its contribution is to identify geometrical measurability as a condition of empirical recoverability: a deeper theory must recover the physical circumstances in which geometrical quantities receive objective determination through stable, recordable, and invariant relations among physical systems. We propose that these conditions concern the stability of the systems that function as measuring devices, the causal accessibility of the systems involved, the production and comparison of records, and the invariance of the relevant quantities under admissible descriptions. In classical general relativity, these requirements may appear almost obvious. In quantum gravity, where the classical infrastructure of geometrical measurement may itself have to emerge from more fundamental degrees of freedom, they become a non-trivial criterion of physical adequacy.

The paper proceeds as follows. Section 2 clarifies the paper’s theoretical positioning. Section 3 defines the physical conditions for objective relational geometrical measurement. Section 4 develops four cases in which these conditions become visible in classical, semiclassical, and limiting geometrical contexts: Rindler horizons and the Unruh effect, black-hole thermodynamics and Jacobson’s equation-of-state derivation, gravitational-wave detection, and Weyl/conformal gravity. Section 5 draws the consequences for quantum gravity and discusses quantum reference frames as a critical domain in which the recovery of objective geometrical determination becomes non-trivial. Finally, Section 6 presents the conclusions.

## 2. Operational roots and theoretical positioning

### *2.1 Operational and chrono-geometrical reconstructions*

Operational reconstructions of spacetime geometry provide the first route to the present account. The Ehlers–Pirani–Schild (EPS) programme reconstructs chrono-geometrical structure from two classes of physical behaviour: the propagation of light and the free fall of massive test particles [9]. Light rays determine the conformal structure by identifying null cones. Freely falling massive particles determine the projective structure by identifying inertial trajectories up to parametrization. Under appropriate compatibility assumptions, these structures allow the recovery of the metric structure characteristic of relativistic spacetime [9,17,18]. Thus, EPS reconstructs geometry rather than replacing it with measurement reports. Its significance lies in giving geometry an empirical genealogy. The metric is introduced through regularities displayed by physically accessible systems such as light signals and freely falling bodies. This supports a modest but important conclusion: the empirical content of spacetime geometry depends on stable physical behaviours that can function as standards of causal and inertial structure [17,18]. A realist about geometry may accept this conclusion, but the conclusion constrains the sense in which metric structure can be treated as physically meaningful independently of the systems through which it is established.

This point is central for the present paper. EPS-style reconstruction shows that the metric is not merely a mathematical field whose empirical interpretation is automatic. The reconstruction requires physically meaningful classes of light rays and freely falling bodies. If those classes are not available, or if their behaviour is not stable enough to function as standards, the geometrical interpretation loses empirical foundation. Thus, even when the final product is a geometrical structure, the route to that structure depends on physical conditions of accessibility, stability, and comparison.

### *2.2 Rods, clocks, and the dynamical meaning of the metric*

A second route comes from the rods-and-clocks problem. In the standard interpretation of relativity, ideal rods measure proper length, and ideal clocks measure proper time. This assumption has physical content. Brown argues that the chrono-geometrical significance of the metric depends on the dynamical behaviour of matter and on the fact that physical rods and clocks track the metric in a universal way [8]. The metric becomes empirically meaningful because material systems behave, under appropriate dynamical conditions, as length- and time-measuring devices.

Historical analyses of Einstein's treatment of rods and clocks sharpen the same issue. Giovanelli [10] shows that Einstein's view moved between treating rods and clocks as primitive measuring instruments and recognizing that their behaviour requires a dynamical account. Darrigol [19] reconstructs the relation between coordinate systems, measurement operations, and the interpretation of the metric in early general relativity. Lehmkuhl [20] argues that Einstein's understanding of general relativity cannot be reduced to the simple claim that gravity was geometrized in the sense of being absorbed into an independently existing spacetime substance.

These analyses shift attention from metric structure alone to the physical conditions under which material systems carry metric information. Rods and clocks are not passive labels attached to a pre-interpreted geometry. They are physical systems whose stability, universality, and dynamical behaviour make proper length and proper time empirically accessible. A theory that does not specify how such systems can function as devices of geometrical determination has not yet recovered the empirical content of metric geometry.

### *2.3 Gauge, observables, and relational individuation*

Diffeomorphism invariance supplies a third route. The hole argument shows that diffeomorphically related models cannot represent distinct physical possibilities merely because fields are assigned differently to bare manifold points [21-23]. The physical content of general relativity cannot be identified with pointwise assignments to an independently labelled manifold. It must be associated with diffeomorphism-invariant structures or relational quantities.

Rovelli's analysis of observables in classical and quantum gravity develops this point in operational terms. In generally covariant systems, observable quantities require physical reference systems and relational localization [24]. The distinction between partial and complete observables clarifies how prediction and measurement can be formulated in systems without an external time parameter [12]. Dittrich further develops this framework for constrained Hamiltonian systems [25].

The relevance for the present argument is direct. Local physical quantities require reference systems; reference systems are themselves physical systems. A coordinate chart is not a physical frame. A physical frame is constituted by material systems, clocks, trajectories, fields, or other physical degrees of freedom whose behaviour enters the relational definition of observable quantities [24,26]. Geometrical objectivity in general relativity is therefore tied to invariant relations among physical systems rather than to bare manifold individuation.

This point fixes the contribution developed here. The paper formulates the physical adequacy conditions under which relational observables become objectively determinable as relations among clocks, rods, signals, detectors, reference systems, and records. Formal gauge invariance secures descriptive admissibility, while empirical access also requires physical instantiation, recordability, and comparison procedures. The physical possibility of this instantiation is the focus of the present paper.

### *2.4 Functionalism and empirical coherence*

Spacetime functionalism provides a fourth route. Knox [27] argues that spacetime structure should be understood through the role it plays in defining inertial frames and organizing the dynamical symmetries of matter. Lam and Wüthrich [15] defend a functionalist approach to spacetime in quantum gravity, according to which spacetime is identified by what it does rather than by an assumed primitive nature. This approach is especially relevant because many quantum-gravity programmes suggest that classical spacetime may be emergent or non-fundamental [13-16,28].

The problem of empirical coherence makes the issue acute. If spacetime is not fundamental, a theory must explain how empirical evidence, apparently gathered in spacetime, can support a theory in which spacetime is emergent or derivative [14,15]. The present study approaches this problem through the physical conditions of empirical access. Empirical coherence is secured when a deeper theory recovers the conditions through which observers, detectors, records, and reference systems instantiate stable spacetime relations. The relevant issue is the physical availability of inertial structure, causal order, metric relations, and curvature effects as measurable and determinable quantities for systems capable of producing comparable records.

### *2.5 Quantum reference frames*

Quantum reference frames sharpen the status of physical reference systems. In classical general relativity, reference frames are often idealized as rods, clocks, congruences, or coordinate systems. In a quantum context, any concrete reference frame is represented by physical systems that may themselves possess quantum properties [11,29]. Indeed, as noted in [29], spacetime descriptions can depend on the quantum system chosen as a reference frame and that proper times associated with different systems need not be related by a single classical transformation.

This changes the epistemological status of localization, duration, and causal description. The systems through which spacetime quantities are defined are part of the physical relational structure. They are not external supports for a pre-given description. Quantum reference frames therefore radicalize a lesson already present in general relativity: objective geometrical determination depends on the physical systems through which relations are defined, compared, and recorded.

For the present paper, quantum reference frames are not merely an additional topic in the foundations of quantum gravity. They are a critical domain in which the proposed conditions become non-trivial. If reference frames are quantum systems, then the emergence of stable clocks, definite records, and mutually comparable relational quantities should no longer be assumed.

## 3. Physical conditions for objective relational geometrical measurement

### *3.1 Definition*

We define the physical conditions for objective relational geometrical measurement as the set of requirements that physical devices, reference systems, detectors, clocks, signals, and records must satisfy in order to yield reproducible and invariant determinations of geometrical quantities. These conditions do not describe an additional dynamical regime of general relativity. They make explicit what is normally implicit in the empirical use of the theory: geometrical quantities become physically meaningful only when they can be instantiated, compared, recorded, and transformed through admissible physical systems.

The term "objective" is used here in a restricted and operationally grounded sense. A geometrical quantity is objectively determined when its value is not a merely subjective assignment, nor an artifact of an arbitrary coordinate description, but a reproducible relational quantity supported by physical systems and invariant comparison procedures.

Four conditions are central.

- First, objective geometrical measurement requires *dynamical stability of physical devices*. Some physical systems must be capable of functioning as clocks, rods, freely falling bodies, mirrors, optical cavities, detectors, or material standards. Their behaviour must be sufficiently robust to support repeatable comparisons of duration, length, phase, trajectory, coincidence, or curvature-related effects [8,10,12]. In classical general relativity, this requirement is often idealized through ideal clocks, ideal rods, test particles, and light signals. The idealization is legitimate only insofar as there exist physical systems whose behaviour approximates these roles.
- Second, objective geometrical measurement requires *causal accessibility*. There must be physical channels through which systems can exchange signals, establish correlations, or influence detector responses. Light rays, particle trajectories, gravitational radiation, and detector couplings define the conditions under which physical systems can access one another [1,9]. Without causal accessibility, geometrical relations may be mathematically definable but physically inaccessible.
- Third, objective geometrical measurement requires *recordability*. Physical interactions must leave "traces" conceptualized and read out as signals that can in principle be compared, stored, transmitted, or embedded in further correlations. Detector readouts, clock records, phase shifts, memory effects, environmental correlations, and persistent relational changes are not secondary to empirical objectivity; they are part of the physical infrastructure through which objectivity is achieved [30,31]. In a theory in which geometry is emergent or relational, the recovery of records is as important as the recovery of the metric itself.
- Fourth, objective geometrical measurement requires *invariance under admissible descriptions*. The relevant quantities must be expressible in relational or gauge-invariant terms. Coordinate-dependent field components can be useful representational tools, but empirical content is secured through invariant structures, complete observables, proper-time relations, detector responses, or relational quantities [12,21-24]. Invariance here does not mean independence from all physical reference systems. It means independence from arbitrary descriptive redundancy once the relevant physical systems have been specified.

Together, these four conditions define what it means for geometrical relations to be objectively determinable. They specify the physical requirements under which the relations encoded by the metric in general relativity can be physically accessible, recordable, and stable under changes of admissible description.

Let us stress that, in classical general relativity, these conditions may appear almost obvious. The theory is normally applied in contexts where rods, clocks, detectors, light signals, freely falling bodies, and records are already presupposed. The metric encodes proper time, causal structure, geodesic motion, curvature, and the behaviour of physical systems used as measurement standards. Its empirical interpretation is therefore supported by the stability of those systems and by a long history of successful measurements, including gravitational redshift, light deflection, orbital dynamics, binary-pulsar timing, gravitational lensing, and gravitational-wave detection. In this classical setting, the four conditions are not additional postulates. They make explicit the physical background that allows the theory to be empirically applied. This apparent obviousness should not be mistaken for triviality. A condition can be obvious in a mature classical theory and non-trivial in a deeper theory that must explain the classical theory's domain of validity. In thermodynamics, for example, the existence of temperature, pressure, and entropy as macroscopic quantities is obvious within the thermodynamic description, but highly non-trivial from the standpoint of microscopic statistical mechanics. Similarly, in classical general relativity the existence of clocks, rods, reference systems, detector records, and gauge-invariant observables is part of the empirical practice of the theory. In quantum gravity, however, these structures may have to emerge from more fundamental degrees of freedom.

This is the main reason for formulating the conditions explicitly. The target is to identify what any deeper theory must recover in order to reproduce the empirical content of general relativity. According to this framework, the recovery of an effective metric or Einstein-like dynamics must be accompanied by the recovery of stable measuring devices, accessible relations, records, and invariant quantities.

### *3.2 Quantum measurement, recordability, and geometrical objectivity*

The notion of objective geometrical measurement used in this paper should also be distinguished from the broader use of "measurement regime" in quantum-measurement theory. In quantum measurement theory, different regimes are usually characterized by the strength, invasiveness, temporal structure, and information gain of measurement interactions. Standard examples include projective measurements, weak and semi-weak measurements, continuous monitoring, generalized von Neumann measurements, and measurement protocols used in discussions of macrorealism, Leggett–Garg inequalities, and no-signaling-in-time conditions [32-37]. These notions are not the primary object of the present analysis. The present account is not concerned with classifying measurement interactions according to their quantum-mechanical strength or back-action, but with identifying the physical conditions under which geometrical relations become empirically accessible, stable, recordable, and invariant in relativistic and gravitational contexts.

This distinction is important because the quantum-measurement problem nevertheless enters the present argument in a more fundamental way. If clocks, rods, detectors, reference frames, and records are physical systems, then in a quantum setting they cannot simply be assumed to possess classical definiteness, stability, or observer-independent outcomes. Their capacity to function as measuring devices depends on dynamical processes through which some correlations become sufficiently robust, persistent, and accessible to support comparison across systems. In this sense, recordability is not merely an epistemic requirement added from outside the theory. It is a physical requirement concerning the emergence of stable traces from quantum interactions.

A record, in the sense relevant here, is not simply a subjective observation or a formal measurement outcome. It is a physical correlation that can persist, be compared, be communicated, or be embedded in further correlations. Detector readouts, clock records, phase shifts, environmental imprints, and persistent relational changes become objective only when they are stabilized enough to support intersubjective and inter-system comparison. This connects with other approaches where objectivity is associated with the emergence of stable and redundantly accessible information from quantum processes [30,31].

This quantum side of recordability strengthens the relevance of the proposed conditions for quantum gravity. A candidate theory must explain the emergence of a classical metric, causal structure, or spacetime geometry together with the emergence of quantum systems sufficiently classical, stable, and correlated to function as clocks, detectors, reference frames, and records. Without such a mechanism, geometrical quantities may be formally definable, but their empirical objectivity would remain incomplete. The recovery of general relativity in a quantum-gravity framework therefore requires a double emergence: effective geometrical structure and the physical measurement infrastructure through which that structure becomes objectively determinable.

This clarification also fixes the intended meaning of the terminology used in this paper. The “conditions” discussed here do not refer to projectiveness, weakness, invasiveness, or measurement back-action in the technical sense of quantum-measurement theory. They refer instead to the stabilization of geometrical measurability: the physical possibility that relational quantities such as proper time, distance, phase, causal accessibility, curvature effects, or detector response become stable enough to be recorded and compared. Quantum measurement theory is relevant insofar as it shows that the existence of such records cannot be treated as primitive in a fundamental quantum description. It must itself be part of what a deeper theory recovers in the appropriate classical limit.

### *3.3 Realism and relation with other approaches*

The position developed in this study may be called a conditional realism about operational geometry. It is realist because gravitational phenomena are treated as physically real when they produce stable, invariant, and reproducible effects. For example, a gravitational redshift is real because clock comparisons yield systematic relations predicted by the metric. According to our framework, a gravitational wave is also real because interferometers register gauge-invariant proper-time and phase effects. In the same sense, an event horizon is real because it structures causal access, thermodynamic bookkeeping, and information recovery for appropriate classes of observers [38-41].

This realism is directed toward invariant physical relations and stabilized access structures. The metric may be fundamental, emergent, thermodynamic, relational, or effective; general relativity by itself does not settle that question [13-15,42]. What general relativity establishes is that gravitational phenomena are represented with extraordinary success by a dynamical geometrical structure whose empirical content is realized through physical systems of measurement, comparison, and record formation.

According to the proposed framework, the conditions should be intended as objective physical conditions rather than merely epistemic preferences. The metric is objectively meaningful because it encodes stable relational determinations supported by physical systems. Whether this structure is fundamental or emergent remains a further question. Their theoretical foundation mostly relies on relational observables and physical reference frames. Indeed, Rovelli’s analysis of observables in classical and quantum gravity shows that local observables in a generally covariant theory require physical reference systems and relational individuation [11]. His distinction between partial and complete observables further clarifies how prediction can be formulated without relying on an external time parameter [12]. Dittrich [25] develops the formal construction of complete observables in constrained Hamiltonian systems. Quantum-reference-frame research radicalizes this point by treating reference frames as quantum systems rather than as external classical structures [29].

On this basis, the present study shifts the emphasis from the formal construction of relational observables to the physical adequacy conditions that make such observables empirically objective. Beside formal gauge invariance, relational quantities require stable devices, causal channels, records, and comparison procedures. A complete observable that cannot be physically instantiated, recorded, or compared remains formally well-defined but empirically incomplete. Conversely, a detector response that is physically recorded but not expressible in invariant terms lacks the objectivity required by general relativity.

This distinction is especially important in quantum gravity. In a quantum-reference-frame setting, the frame itself may be indefinite, entangled, or subject to quantum transformations. A quantum-gravity theory must therefore account for the transformation of descriptions between quantum frames and for the stabilization of relational quantities that support objective comparison and record formation. The four conditions proposed here articulate this requirement without selecting a specific quantum-gravity programme.

As discussed in Section 2, the proposal is grounded in different epistemological approaches to gravity and overlaps with spacetime functionalism. From a functionalist perspective, spacetime is identified by its role in the field equations and by the physical conditions through which that role becomes empirically available. Functionalism can explain why an emergent structure counts as spacetime if it plays the appropriate spacetime roles, especially in defining inertial structure and organizing matter dynamics [15,27]. The present proposal adds that those roles must be physically accessible to systems that can function as observers, detectors, or records. A structure that plays the formal role of a metric in an effective equation must also support stable clocks, accessible signals, records, and invariant comparison procedures in order to recover spacetime as an empirical structure. This yields a sharpened epistemological constraint: general relativity teaches quantum gravity to recover the physical possibility of objective geometrical determination together with geometry itself.

## 4. Classical, semiclassical, and limiting cases of objective geometrical determination

The cases we will discuss in this section do not all have the same theoretical status. Some are internal to classical general relativity, some belong to semiclassical or quantum-theoretical regimes at the boundary of general relativity, and some concern alternative or extended geometrical frameworks. Their common function is not to test general relativity from outside, but to make explicit where the physical conditions of objective geometrical determination become visible, non-trivial, or potentially restrictive.

### *4.1 Rindler horizons and the Unruh effect: causal access and detector response*

Rindler horizons provide a controlled case in which horizon-like structure arises without spacetime curvature. A uniformly accelerated observer in Minkowski spacetime has access only to a Rindler wedge. The boundary of that wedge functions as a horizon for that observer, although the underlying spacetime is flat [21,43,44]. The horizon is not a curvature singularity or an absolute boundary of the full spacetime. It is a structure of causal accessibility associated with a physically specifiable class of trajectories.

The Unruh effect deepens this point. An accelerated detector in the Minkowski vacuum responds as if immersed in a thermal bath with temperature proportional to its proper acceleration [44]. Algebraic analyses, including the Bisognano–Wichmann theorem and Sewell's work on quantum fields on manifolds, connect the thermal character of restricted vacuum states to wedge localization and modular structure in quantum field theory [43,45]. Earman [21] emphasizes that the Unruh effect has multiple formulations, including detector response, algebraic thermality, and particle interpretations. Its conceptual importance lies in connecting acceleration, localization, detector response, thermality, and horizon structure.

From the standpoint of the proposed conditions, the Rindler–Unruh case is instructive because it satisfies them in a minimal and controlled form. The accelerated detector provides the physical device. Its worldline defines the relevant reference system. The Rindler wedge specifies causal accessibility. The detector response provides a recordable physical effect. The relevant temperature and response functions are not arbitrary coordinate artifacts, but are associated with invariant features of the accelerated trajectory and the field state restricted to the accessible wedge.

This case shows how a geometrical and thermodynamic structure acquires empirical meaning only when there exists a physical system capable of transforming causal accessibility into measurable response. The Rindler horizon is observer-dependent, but it is not subjective. Its observer-dependence is tied to a physically specified class of accelerated systems. The Unruh response is likewise not a mere coordinate description. It is a calculable detector effect. Thus, the Rindler–Unruh case demonstrates how causal access, device dynamics, recordability, and invariance jointly contribute to objective relational geometrical determination.

The lesson is limited but important. A horizon-like structure can be physically meaningful even in flat spacetime when the relevant physical systems and access conditions are specified. This lesson becomes more powerful when combined with black-hole thermodynamics, where horizons are not merely acceleration horizons in Minkowski spacetime but central features of gravitational dynamics.

### *4.2 Black-hole thermodynamics and Jacobson: geometry, entropy, and macroscopic consistency*

Black-hole thermodynamics links spacetime geometry to entropy, temperature, causal horizons, and information. Bekenstein [38] argued that black holes carry entropy proportional to horizon area. Hawking [39] showed that quantum fields on black-hole backgrounds lead to thermal radiation with temperature proportional to surface gravity. Wald [46] later showed that, in a broad class of diffeomorphism-invariant theories, black-hole entropy can be associated with a Noether charge. Later reviews establish black-hole thermodynamics as a central interface between general relativity, quantum field theory, statistical mechanics, and information theory [2,47].

Entanglement entropy reinforces the connection between geometry, access, and correlations. Sorkin's early work identified the entropy of quantum fields outside a horizon as a candidate source of black-hole entropy, with an area-law behaviour for correlations across a boundary [48]. Solodukhin [49] reviews the role of entanglement entropy in black-hole physics and its dependence on correlations between degrees of freedom separated by a surface. Bianchi [50] links horizon entanglement entropy to the universality of gravitational coupling. These results show that horizon entropy is tied to partitions of access, correlations across boundaries, and the geometrical structure of causal separation [48-50].

Jacobson's equation-of-state derivation gives this connection a particularly sharp form. Jacobson [42] derives the Einstein equation from the Clausius relation, the proportionality of entropy to horizon area, and the Unruh temperature associated with local Rindler horizons. The Einstein equation is interpreted as an equation of state, analogous to macroscopic thermodynamic or hydrodynamic relations. Jacobson and Parentani [51] extend the significance of horizon entropy by associating entropy with causal horizons more generally.

From the perspective developed here, Jacobson's derivation is an example in which gravitational dynamics is recovered as a macroscopic consistency condition among local causal horizons, energy flux, entropy, and acceleration temperature. The derivation therefore illustrates how geometrical dynamics can be constrained by physical conditions of accessibility, thermality, and recordability. Local Rindler horizons provide the causal-access structure; energy flux across the horizon supplies the relevant physical interaction; entropy variation functions as a macroscopic bookkeeping variable for inaccessible degrees of freedom; the Clausius relation imposes a consistency condition; and the resulting Einstein equation expresses the geometrical dynamics compatible with these conditions.

The black-hole thermodynamic case therefore supports the general claim that the empirical content of geometry is not exhausted by the formal existence of a metric. Geometry, thermality, causal access, and information appear as mutually connected aspects of the physical conditions under which gravitational dynamics acquires meaning. The metric may function analogously to a macroscopic state variable: real and indispensable at its level, while open to deeper explanation.

Finally, this does not commit the paper to a thermodynamic or emergent-gravity ontology. The same case can be interpreted differently by different approaches. Its relevance lies in showing that, even within semiclassical and thermodynamic reasoning, the recovery of gravitational dynamics is tied to the recovery of physical structures of access, correlation, entropy, and invariant comparison.

### *4.3 Gravitational waves: detector-accessible geometrical reality*

Gravitational waves are a decisive case because they strongly support realism about dynamical spacetime structure. The first LIGO detection confirmed a central prediction of general relativity and established gravitational-wave astronomy as an observational field [4]. Advanced interferometric detectors are designed to detect extremely small phase and effective arm-length differences induced by passing gravitational radiation [3]. These observations show that gravitational waves are real physical phenomena.

Their empirical access, however, is mediated by gauge-invariant detector observables. A detector does not measure a bare component of a metric perturbation. Metric perturbations are gauge-dependent, while detector observables must be formulated in gauge-invariant terms [40,41,52]. Koop and Finn [40] provide a gauge-invariant description of light-time detector response. Błaut [52] derives the response of a laser interferometer to a plane gravitational wave without relying on a particular gauge within the relevant class. Lee and Zurek [41] formulate proper-time observables for laser-interferometry-based gravitational-wave detectors, identifying the relevant observable with the proper time elapsed along the beamsplitter observer's worldline between photon departure and return.

Gravitational-wave detection satisfies the four proposed conditions in a particularly transparent way. The detector is dynamically stabilized through mirrors, lasers, optical cavities, suspensions, timing systems, and control systems. The wave is causally accessible through its coupling to optical paths and test masses. The signal is recordable as a phase or proper-time observable. The measured quantity is not a coordinate-dependent metric perturbation, but a detector response that can be expressed in gauge-invariant relational terms.

Persistent gravitational-wave observables support the same reading. Flanagan et al. [53] develop a general framework for persistent observables, including gravitational-wave memory. Grant [54] extends the analysis of nonlinear persistent observables in curved spacetime. These effects are physically significant because they correspond to lasting changes in relations among observers, worldlines, or detector components [53,54]. They show how dynamical spacetime structure becomes empirically accessible through invariant relational change.

The gravitational-wave case clarifies the realist content of the present account. Gravitational waves are real because they produce reproducible, gauge-invariant, detector-accessible effects. Their reality is expressed geometrically, but their empirical objectivity is secured through physical systems of measurement. This is why the case is especially important for quantum gravity. A deeper theory that recovers gravitational waves only as perturbations of an effective metric, without recovering the detector-accessible invariant quantities through which they become empirical, would recover the formal structure but not the full epistemic content of the phenomenon.

### *4.4 Weyl and conformal gravity as a limiting case for geometrical measurability*

Weyl and conformal gravity provide a critical limiting case for the proposed conditions. The aim of this section is to study whether the conditions of objective relational geometrical measurement can be satisfied in a theory in which conformal invariance changes the status of scale-dependent quantities such as length, duration, and curvature scales.

The issue is conceptually sharp. Under a Weyl transformation, the metric is rescaled locally, typically as $g_{\mu\nu} \to \Omega^2(x) g_{\mu\nu}$. If the theory is genuinely invariant under such transformations, then the conformal class rather than a single metric may carry the fundamental geometrical information. The causal structure associated with null cones is preserved, but scale-dependent quantities such as proper time, proper length, and curvature scalars can change unless a physical mechanism fixes the conformal frame or makes the relevant quantities conformally invariant. This raises an immediate question for the present account: can a conformally invariant gravitational theory provide objective determinations of space and time measurements without an additional physical prescription for rods, clocks, and scale-setting matter?

This question is not merely philosophical. Conformal and Weyl-invariant theories are mathematically rich and have been explored both as alternatives to general relativity and as possible ingredients in quantum-gravity-inspired models. Mannheim and Kazanas [55] developed a conformal-gravity solution with phenomenological relevance for galactic rotation curves. Mannheim [56] reviews conformal gravity as a possible alternative framework. Rachwał [57] discusses quantization issues for conformal gravity. Bambi, Modesto, and Rachwał [58] argued that Weyl conformal symmetry can lead to geodesically complete, non-singular black-hole spacetimes conformally related to Schwarzschild or Kerr geometries. Modesto, Zhu, and Zhang [59] argue that the Kasner spacetime can be made singularity-free in Einstein conformal gravity. Cadoni et al. [60] construct Lorentzian traversable wormholes in conformal gravity through Weyl rescalings of Minkowski spacetime, arguing that such wormholes are solutions of any conformally invariant gravitational theory admitting Minkowski spacetime as a solution.

These results are important for the present discussion because they show how physically distinct-looking geometries may be related by conformal transformations. In particular, if Minkowski spacetime can be Weyl-rescaled into a wormhole geometry within a conformally invariant framework, the question of what is physically measured becomes delicate. Are the two descriptions physically equivalent, different phases, different gauge choices, or physically distinguishable only after conformal symmetry breaking or matter coupling selects a measurement scale? If a singular geometry can be transformed into a regular one through a Weyl rescaling, what is the operational status of the singularity? Does the transformation remove a physical pathology, or does it move the issue into the interpretation of rods, clocks, and conformal frame selection?

From the standpoint of the four conditions, conformal gravity poses the following problem. The causal-accessibility condition may be partly preserved, because conformal transformations preserve null structure. However, the metric-stabilization condition becomes non-trivial, because proper time and proper length are not fixed by the conformal structure alone. The dynamical-stability condition also becomes non-trivial, because rods and clocks must be associated with matter dynamics, symmetry breaking, or conformally invariant measurement prescriptions. The recordability condition requires that detector records be expressed in quantities that remain physically meaningful under conformal transformations or that a physical conformal frame be selected. The invariance condition requires distinguishing between gauge redundancy and physically different configurations.

This does not imply that Weyl or conformal gravity necessarily fails the proposed conditions. It implies that the satisfaction of the conditions is not automatic. A conformally invariant theory may satisfy them if it provides a physical mechanism that fixes the relevant scale, such as spontaneous symmetry breaking, matter coupling, boundary conditions, or conformally invariant definitions of observables. Conversely, if the theory permits transformations that erase or introduce length and time determinations without specifying how physical devices distinguish equivalent from inequivalent configurations, then it risks failing the metric-stabilization condition and may yield geometrical predictions whose observational content is underdetermined.

The case of conformally generated wormholes is especially instructive. If a wormhole geometry is obtained by Weyl-rescaling Minkowski spacetime, and if the theory treats conformally related metrics as gauge-equivalent prior to symmetry breaking, then one must ask what physical device would measure the wormhole throat, the traversal time, the curvature scale, or the proper distance. If the answer depends on a selected conformal frame, then the selection of that frame must be physically justified. If the answer is conformally invariant, then the theory must identify the invariant quantities corresponding to the claimed observable effects. Without such a specification, the geometry may be mathematically well-defined but not yet empirically determinate.

The same issue appears in conformal treatments of singularities. If singularities are removed by Weyl rescalings, this may represent an important physical mechanism. However, from the present standpoint, the decisive question is whether the removal is accompanied by a physical account of probes, clocks, affine parameters, records, and invariant observables. A regular conformal representative is not by itself enough to establish observational regularity unless the physical measurement prescriptions are also specified.

For this reason, Weyl and conformal gravity should be treated as a future research direction rather than as a settled conclusion of the present paper. The present analysis suggests a concrete programme: evaluate conformal-gravity models by asking whether they satisfy the four conditions of objective relational geometrical measurement. Such an analysis would have to examine the matter sector, symmetry breaking, conformal-frame selection, clock and rod behaviour, detector observables, and the operational meaning of conformally related geometries. The possible outcome is open. Some conformal models may satisfy the conditions once the physical frame is fixed. Others may fail to produce determinate predictions for space and time measurements unless additional structure is introduced.

## 5. Consequences for quantum gravity

The cases discussed in the previous section show how the four conditions of objective relational geometrical measurement operate across classical, semiclassical, and limiting geometrical contexts. Rindler horizons and the Unruh effect make explicit the role of causal accessibility and detector response. Black-hole thermodynamics and Jacobson's equation-of-state derivation connect geometry with entropy, horizon structure, and macroscopic consistency. Gravitational-wave detection shows that dynamical spacetime geometry becomes empirical through gauge-invariant detector observables. Weyl and conformal gravity provide a limiting case in which causal structure may be preserved while scale-dependent measurements of length, duration, and curvature require additional physical specification.

**Table 1.** Physical conditions for objective geometrical determination across the analysed cases.

| Case | Dynamical stability | Causal accessibility | Recordability | Invariance |
|---|---|---|---|---|
| Rindler/Unruh | accelerated detector | Rindler wedge | detector response | trajectory-dependent invariant response |
| Black-hole thermodynamics/Jacobson | local horizon systems | causal horizons | entropy/energy flux | thermodynamic-geometrical consistency |
| Gravitational waves | interferometer infrastructure | coupling to optical paths/test masses | phase/proper-time signal | gauge-invariant detector observable |
| Weyl/conformal gravity | scale-setting matter/rods/clocks | conformal causal structure | conformally meaningful detector records | frame/gauge distinction under Weyl transformations |

Table 1 condenses the role of the four conditions across the analysed cases. Taken together, these cases show that objective geometrical determination depends on the coordinated operation of stable physical systems, causal channels, record-forming interactions, and invariant comparison procedures. Quantum reference frames enter at the next step: they do not simply provide another example of these conditions, but make explicit why their recovery becomes a non-trivial requirement in the transition from quantum descriptions to classical spacetime geometry.

### *5.1 From recovering geometry to recovering objective determination*

The proposed account reframes the epistemological target of quantum gravity. A deeper theory must explain how the structures represented geometrically in general relativity arise and how they become physically available as objective determinations of relational geometrical quantities.

This point matters because the standard motivation for quantum gravity is often compressed into the claim that gravity must be quantized. A deeper theory is certainly required to address the relation between quantum matter, gravitational dynamics, black-hole thermodynamics, cosmology, and the information problem [2,13,14]. However, the inference from the need for a deeper theory to the claim that the classical metric must be quantized as a primitive field is not logically immediate. Huggett and Callender [61] argue that the need to reconcile quantum theory and gravity does not by itself prove that gravity must be quantized in the same sense as other fields. Wüthrich [62] shows that the motivations for quantizing gravity are more subtle than standard folklore suggests. Mattingly [63] and Kent [64] criticize the Eppley–Hannah argument [65] concerning a possible violation of either momentum conservation or the uncertainty principle when a classical gravitational field interacts with a quantum system, resulting in transmission of signals faster than c, which has often been used to motivate the necessity of quantizing gravity.

Recent semiclassical and postquantum proposals keep this logical space open, although they remain controversial and empirically constrained. Tilloy [66] argues that non-relativistic toy models make it difficult to dismiss semiclassical gravity on purely conceptual grounds. Oppenheim [67] proposes a postquantum framework in which classical gravity couples to quantum fields through a linear, completely positive, trace-preserving dynamics. The present account does not depend on the success of such proposals. Their methodological significance is that they prevent the problem of quantum gravity from being reduced too quickly to the direct quantization of a primitive metric field.

The proposed conditions apply across competing approaches. If the correct theory quantizes geometrical degrees of freedom, it must explain how classical geometrical measurement becomes possible from quantum gravitational states, histories, or amplitudes. If spacetime emerges from non-spatiotemporal structures, the theory must explain how observers, detectors, records, and reference systems become embedded in the emergent structure in a way that allows evidence to be gathered and compared [14,15]. If gravity is described by an effective, thermodynamic, or hydrodynamic theory, the theory must explain why pseudo-Riemannian geometry with Lorentzian signature provides the correct macroscopic representation of accessible relations among physical systems [13,42,68].

Recovering general relativity therefore involves at least four epistemic-physical components. First, a theory must recover causal stabilization: stable structures of causal order, signalling, and accessibility through which physical systems can interact and exchange information. Second, it must recover metric stabilization: stable proper-time, distance, phase, and trajectory relations among physical systems capable of functioning as clocks, rods, detectors, or reference systems. Third, it must recover record stabilization: physical traces that can persist long enough to be compared, communicated, or embedded in further correlations. Fourth, it must recover invariance stabilization: relational or gauge-invariant quantities that remain physically meaningful across coordinate choices, reference systems, and admissible descriptions.

These requirements specify a programme-independent criterion for recovering the empirical content of general relativity. A candidate theory must recover an effective metric, a continuum limit, or Einstein-like dynamics together with the physical conditions through which metric relations become measurable, recordable, and shareable. In this sense, quantum gravity is the problem of recovering spacetime and the physical possibility of spacetime measurement.

### *5.2 Emergent and effective gravity: why the conditions become restrictive*

Emergent and effective approaches make the proposed conditions especially restrictive. In these frameworks, gravitational phenomena may arise from entropy, horizons, or macroscopic consistency conditions [28,42,68,69], from collective or emergent degrees of freedom [70-74], from analogue effective metrics [75,76], or from modified localization structures in relative-locality and curved-momentum-space approaches [77-82]. Across these domains, the relevant question is how the emergent or effective structure supplies stable devices, causal or effective causal channels, record-like correlations, and invariant relational quantities.

The constraint is therefore constructive: it specifies what has to emerge together with an effective metric or Einstein-like equation. Geometry must be accompanied by the physical conditions under which it can be measured, recorded, and invariantly compared.

### *5.3 Quantum reference frames and the recovery of classical objectivity*

Quantum reference frames make this requirement especially sharp. In classical general relativity, reference frames are often idealized as rods, clocks, congruences, or coordinate systems. This idealization is acceptable when the physical systems functioning as frames can be treated as sufficiently stable, classical, and external to the measurement being described. In a quantum context, this assumption is no longer harmless. Reference frames are physical systems, and physical systems may themselves be quantum.

Rovelli's analysis of quantum reference systems already showed that reference systems should not be treated as external absolute structures [11,24]. Giacomini's work on spacetime quantum reference frames develops this point by showing that spacetime descriptions can depend on the quantum system chosen as a reference frame and that proper times associated with different systems need not be related by a single classical transformation [29]. Localization, simultaneity, duration, and causal description can therefore become relational in a stronger sense than in classical general relativity.

This does not make geometrical quantities merely subjective or arbitrary. A quantity may be relational to a physical quantum frame and still be objective, provided that the frame, the interaction, the record, and the transformation rules are physically specified. Conversely, a quantity may be formally defined but fail to be objectively determinable if no physical procedure allows it to be stabilized, recorded, or compared across admissible descriptions. Quantum reference frames therefore expose the precise point at which the four conditions proposed in this paper become necessary.

The dynamical-stability condition requires that some quantum systems behave, at least approximately and relationally, as clocks, rods, detectors, or reference frames. The causal-accessibility condition requires physically meaningful correlations or interactions between the systems whose relations are to be compared. The recordability condition requires that relational information become encoded in physical traces capable of supporting comparison across perspectives. The invariance condition requires that predictions not depend on arbitrary descriptive choices, even when they are relational to a chosen quantum frame.

Quantum reference frames therefore clarify the classical limit of quantum gravity. A classical spacetime description should emerge when quantum systems support relational quantities stable enough to reproduce proper-time relations, causal order, detector responses, and invariant comparison procedures. The classical frame is not a primitive background; it is the limiting case in which physical reference systems become sufficiently stable for geometrical determinations to be treated as objective in the classical sense.

This point connects the formal problem of quantum reference frames with the epistemological constraint developed in this paper. A quantum theory must assign states to geometry, transform descriptions between frames, define relational observables, and explain how quantum systems become the physical infrastructure through which geometry is determined. In this sense, quantum reference frames provide an admissibility test for quantum-gravity approaches: they show that recovering general relativity requires recovering the conditions under which frame-dependent quantum descriptions give rise to stable, recordable, and invariant classical geometrical determinations.

## 6. Conclusion

This paper examines the epistemological significance of general relativity for quantum gravity from the standpoint of objective relational geometrical measurement. The starting point was the distinction between treating the metric as a primitive geometrical entity and treating it as the objective representation of stable relations among physical systems. The aim was not to deny the geometrical character of general relativity, but to clarify what makes that geometry empirically meaningful. The central proposal is that general relativity specifies physical conditions under which relational geometrical quantities can be objectively determined: dynamical stability of devices, causal accessibility, recordability, and invariance under admissible descriptions.

The argument was developed in four steps. First, the paper reconstructed several routes that already point toward this interpretation: operational reconstructions of chrono-geometrical structure, dynamical analyses of rods and clocks, diffeomorphism invariance, relational observables, spacetime functionalism, and quantum reference frames. Taken separately, these approaches address different aspects of the empirical meaning of geometry. Taken together, they show that metric structure becomes physically significant through conditions of access, comparison, stabilization, and record formation.

Second, the paper formulated these requirements as conditions for objective relational geometrical measurement. In classical general relativity, these conditions may appear almost obvious because the empirical practice of the theory already presupposes clocks, rods, light signals, freely falling bodies, detectors, reference systems, and records. Their significance becomes non-trivial in quantum gravity, where these structures may need to emerge from more fundamental degrees of freedom.

Third, the paper analysed four cases in which the proposed conditions become especially visible. Rindler horizons and the Unruh effect showed how causal access, acceleration, localization, detector response, and thermality can define physically meaningful horizon-like structures even in flat spacetime. Black-hole thermodynamics and Jacobson's equation-of-state derivation showed how geometry, entropy, horizon structure, and macroscopic consistency are deeply connected. Gravitational-wave detection showed how one of the strongest empirical confirmations of general relativity is secured through gauge-invariant detector observables rather than through direct access to coordinate-dependent metric perturbations. Weyl and conformal gravity were discussed as a critical limiting case: conformal invariance preserves causal structure but can make scale-dependent measurements of length and time non-trivial unless a physical conformal frame, symmetry-breaking mechanism, matter coupling, or conformally invariant observable prescription is specified. Section 5 then used emergent gravity and quantum reference frames to show why these conditions become genuinely restrictive in quantum gravity. If the systems used to define spacetime quantities are themselves quantum, the recovery of stable reference systems, comparable records, and invariant relational quantities becomes part of the classical limit rather than a background assumption.

The argument developed in this paper is ontologically neutral but epistemologically restrictive. It is compatible with interpretations in which spacetime geometry is fundamental, emergent, relational, functional, or effective, while identifying a condition that any such interpretation must satisfy: the geometrical content of general relativity becomes empirically meaningful when physical systems instantiate stable, causally accessible, recordable, and invariant relations. This point is consistent with operational reconstructions of chrono-geometrical structure, dynamical analyses of rods and clocks, relational observables in generally covariant systems, spacetime functionalism, and quantum reference frames. The recovery of general relativity in quantum gravity therefore requires the mathematical form of spacetime geometry and the physical conditions under which geometrical quantities can be objectively determined by physical devices and reference systems.

The broader implication is that general relativity teaches quantum gravity a double lesson. It shows that gravitation can be represented geometrically with extraordinary empirical success. It also shows that this geometrical representation becomes meaningful only when physical systems stabilize access, comparison, records, and invariance. Recovering geometry is therefore not the whole task. A satisfactory theory of quantum gravity must also explain how the world comes to contain the physical conditions under which geometry can be measured.

This perspective opens several directions for future work. A first direction is conceptual: the proposed conditions should be compared more systematically with existing accounts of empirical coherence, spacetime functionalism, relational observables, and quantum reference frames. A second direction is theory-specific: different approaches to quantum gravity should be analysed in terms of how they recover causal stabilization, metric stabilization, record stabilization, and invariance stabilization. Loop quantum gravity, spin-foam models, causal-set theory, group-field theory, asymptotic-safety scenarios, analogue and emergent-gravity models, semiclassical gravity, postquantum proposals, and conformal-gravity scenarios may satisfy these requirements in different ways. A third direction concerns concrete measurement models. Gravitational-wave detectors, atomic clocks in gravitational fields, quantum-clock interferometry, horizon thermodynamics, and quantum-reference-frame scenarios provide natural domains in which the physical infrastructure of geometrical measurability can be analysed in detail. A fourth direction concerns conformal and Weyl-invariant theories. These theories should be examined for their mathematical consistency, phenomenological predictions, and ability to specify what physical devices measure when conformal transformations change scale-dependent geometrical quantities. A fifth and broader direction would be to extend the present framework beyond the recovery of general relativity. The present paper has focused on the empirical meaning of spacetime geometry, and therefore on the conditions under which relational geometrical quantities become objectively determinable. However, a satisfactory quantum-gravity theory should also recover, in the appropriate limits, the field-theoretical and quantum structures through which high-energy physics and ordinary quantum mechanics acquire empirical content. This would require analysing the emergence of classical geometry together with the recovery of stable particle-like excitations, scattering observables, gauge-invariant quantities, detector interactions, preparation procedures, and record-forming processes. In this broader perspective, the conditions proposed here may be viewed as the geometrical-gravitational instance of a more general requirement of empirical recoverability: a deeper theory must recover the formal structures of its effective limits together with the physical conditions under which those structures become measurable, recordable, and invariantly comparable.

## Acknowledgements

The author thanks M. Cadoni for useful discussions and comments that helped improve the quality of the paper.

### Funding

There has been no significant financial support for this work that could have influenced its outcome.

### Competing interest

There are no known conflicts of interest associated with this publication.

**Declaration of generative AI and AI-assisted technologies in the writing process**

During the preparation of this manuscript, the authors used ChatGPT to support language revision, citation-style harmonisation, and editorial restructuring of selected sections. After using this tool, the authors reviewed and edited the content as needed and take full responsibility for the content of the publication. Generative AI tools were not used to design the study, collect the data, code the corpus, interpret the results, or create or alter the figures included in the manuscript.

**Orcid ID**

Matteo Tuveri https://orcid.org/0000-0001-5686-1713